\documentclass[%
 reprint,
superscriptaddress,
 amsmath,amssymb,
prstab,
]{revtex4-2}

\usepackage{caption}
\usepackage[caption=false]{subfig}
\usepackage{graphicx}
\usepackage{dcolumn}
\usepackage{bm}
\usepackage{color}
\usepackage{booktabs}

\begin{document}
\preprint{APS/123-QED}

\title {Nonlinearly Shaped Pulses in Photoinjectors and Free-Electron Lasers}

\author{Nicole Neveu}\thanks{Corresponding author: nneveu@slac.stanford.edu}
\affiliation{SLAC National Accelerator Laboratory, Menlo Park, CA, USA}
\author{Randy Lemons}
\affiliation{SLAC National Accelerator Laboratory}
\affiliation{Colorado School of Mines, Golden, CO, USA}
\author{Joseph Duris}
\author{Jingyi Tang}
\affiliation{SLAC National Accelerator Laboratory}
\author{Yuantao Ding}
\affiliation{SLAC National Accelerator Laboratory}
\author{Agostino Marinelli}
\affiliation{SLAC National Accelerator Laboratory}
\author{Sergio Carbajo}
\affiliation{SLAC National Accelerator Laboratory}
\affiliation{University of California, Los Angeles, CA, USA}

\date{\today}

\begin{abstract}

Photoinjectors and  Free Electron Lasers (FEL) are amongst the most advanced systems in accelerator physics and have consistently pushed the boundaries of emittance and x-ray peak power.
In this paper, laser shaping at the cathode is proposed to further lower the emittance and reduce electron beam tails, which would result in brighter x-ray production. Using dispersion controlled nonlinear shaping (DCNS), laser pulses and beam dynamics were simulated in LCLS-II.
The photoinjector emittance was optimized and the resulting e-beam profiles were then simulated and optimized in the linac.
Finally, the expected FEL performance is estimated and compared to the current technology: Gaussian laser pulses on the cathode.
The e-beams produced by DCNS pulses show a potential for 35\% increase in x-ray power per pulse during SASE when compared to the standard Gaussian laser pulses.
\end{abstract}

\maketitle

\section{\label{sec:intro}Introduction}
Accelerator performance and optimization is the foundation of advances in state-of-the-art ultrafast and fundamental space-time resolution instrumentation.
Advances in light sources, emerging medical, industrial, and ultrafast electron diffraction applications would not be possible without pushing the boundaries of the underlying accelerator physics. From free electron generation to secondary radiation emission, improvements to the photoinjector, linear accelerator (linac) and undulator technologies and methodologies are being explored with the goal of producing precisely controlled electron beams and photon pulses~\cite{PhysRevLett.123.214801,PhysRevLett.124.074801,PhysRevLett.125.044801, PhysRevLett.126.104802}.

In this work, reaching the next level of performance (lower emittances and higher x-ray pulse energies) in photoinjectors and X-ray Free Electron Laser (XFEL) facilities is explored through laser shaping. As the first step of a long cascade of processes, the laser shape on a photocathode sets many performance boundaries observed down stream in accelerators, including emittance and bunch length. Photoinjector pulse shaping may be a foundational approach to tackling these limits in accelerator-based (ultrafast) instrumentation relying on photoemisison, and we consider the temporal laser shape at the photocathode as a means to improve the emittance and longitudinal profile of the beam.

\begin{figure}
    \includegraphics[width=0.95\linewidth]{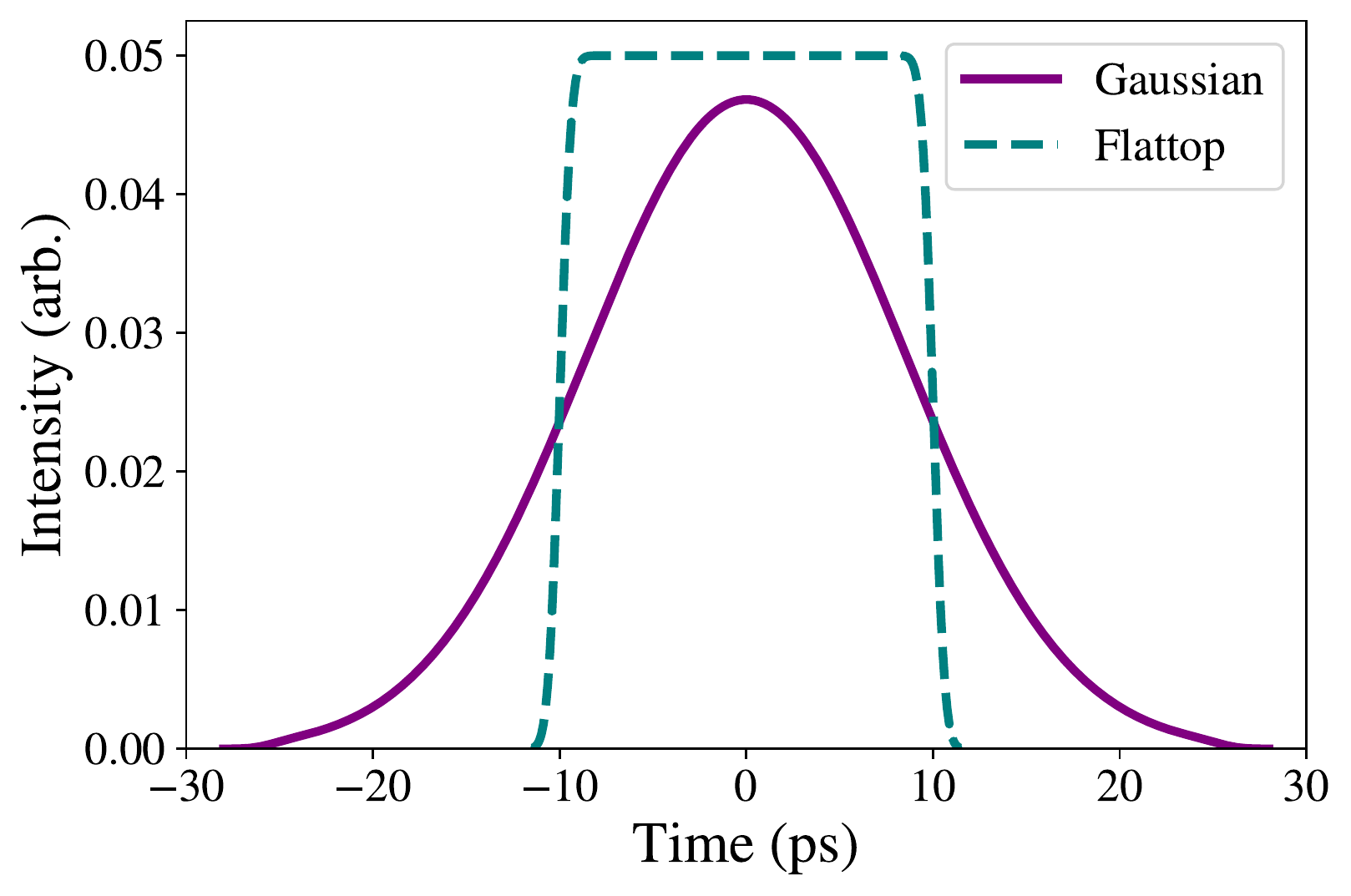}
    \caption{Comparison of Gaussian and flattop laser pulses used in simulation. The Gaussian profile matches what is currently generated at the LCLS-II photoinjector laser room, and is being used for commissioning.}
    \label{fig:flatvsgauss}
\end{figure}

Currently, the operational ultraviolet (UV) optics installed at the Linac Coherent Light Source (LCLS) and LCLS-II produce temporally Gaussian-shaped pulses~\cite{gilevich2020lcls}. Previous research~\cite{krasilnikov2012experimentally,krasilnikov2019studies,petrarca2007production}, indicates a flattop UV laser pulse at the cathode is ideal. Typical longitudinal laser pulse shapes used in studies for LCLS-II are shown in Fig.~\ref{fig:flatvsgauss}. Simulation studies for LCLS-II show Gaussian pulse shapes, also shown in Fig.~\ref{fig:flatvsgauss}, result in larger emittances when compared to the ideal flattop pulses used in design studies (per a given bunch length)~\cite{technote1}.  Despite the prediction of lower performance, the repetition rate of 1 MHz is achievable, and the Gaussian UV pulses will be used for initial LCLS-II commissioning.

Flattop UV pulses have been achieved through methods {such as spatial-light modulators~\cite{mironov2016shaping,penco2013optimization}, acousto-optic modulators~\cite{li2009laser,petrarca2007production}, and pulse stackers~\cite{krasilnikov2012experimentally,krasilnikov2019studies}}, but there remains a technical void and lack of techniques that can produce high-fidelity, efficient, high-quality flattop UV pulses like the ones used in simulation studies. Moreover, a flattop laser profile has not been demonstrated at high per-pulse energy and the unprecedentedly high 1~MHz repetition rate for the high average powers needed for machines like the LCLS-II.
In an effort to approach or reach the ideal flattop performance, alternative optics methods and pulse shapes are being investigated and motivate this work.

\section{Optics}\label{sec:optics}
The commissioning LCLS-II photoinjector optical design requires 257 nm wavelength pulses, a Gaussian temporal shape with 20 ps duration at full width half maximum (FWHM), and ~0.5 $\mu$J pulse energy. Currently, the front-end optical driver employs an infrared (IR) Ytterbium-based laser amplifier operating at 50 $\mu$J energy at 1 MHz repetition rate, and 330 fs compressed pulse duration at 1030 nm central wavelength~\cite{gilevich2020lcls}. The IR pulses are first upconverted to the deep UV through two subsequent second-harmonic generation (SHG) steps in $\beta$-Barium Borate (BBO) crystals. Once converted, the UV pulses are dispersed to 20 ps FWHM using a single-grating pulse stretcher. The total efficiency of the conversion and stretching is $\approx$10\%, thereby leaving ~5W of usable UV power to transport to the photocathode.

\begin{figure}[htp!]
    \includegraphics[width=0.95\linewidth]{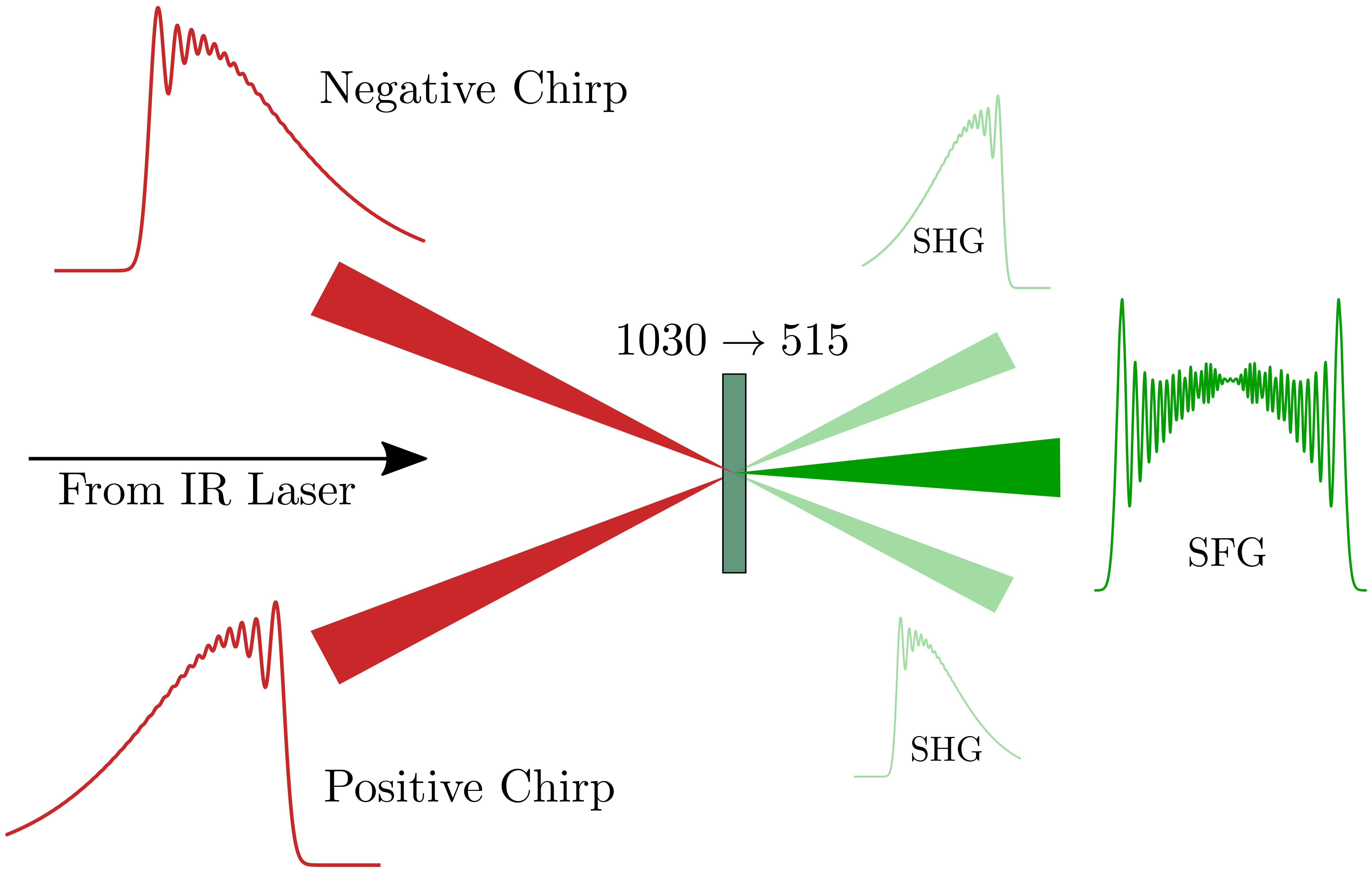}
    \caption{Sum-frequency generation of a long optical pulse with shaped temporal intensity profile from highly dispersed short pulses. Temporally flattop pulses at 515 nm are generated in a non-colinear geometry from 1030 nm pulses with tailored phase}
    \label{fig:SFG_Only_OpChirp}
\end{figure}

To generate UV pulses with flattop temporal shapes, we employ a cascaded nonlinear conversion chain where the DCNS technique is applied to the first nonlinear conversion stage (1030 nm $\rightarrow$ 515 nm). Conceptually, this DCNS implementation consists of splitting a fundamental broadband IR pulse into two arms, then applying equal and opposite amounts of second-order dispersion (SOD) and third-order dispersion (TOD) and mixing them in a non-colinear sum-frequency generation (SFG) scheme~\cite{lemons2022temporal}. The process results in a frequency-doubled pulse whose bandwidth is greatly reduced relative to a colinear SHG scheme. To achieve a temporally flat-top and nearly transform-limited upconverted pulse, the ratio of SOD and TOD imposed onto the IR pulses must yield two temporally semi-triangular pulses, as seen in Fig. \ref{fig:SFG_Only_OpChirp}. This technique has the added benefit of efficiently producing SFG pulses with a flattened spectral phase, thereby circumventing phase-induced temporal distortions in further upconversion stages.

For the current study, we begin by modeling the mixed IR pulses as a combination of two equal-energy transform-limited Gaussian pulses overlapped in time. From here, the phase of each pulse can be adjusted separately by multiplying by $e^{i \varphi(\omega)}$ in frequency space where $\varphi(\omega)$ is given by

\begin{multline}
    \varphi(\omega) = \varphi_0 + \varphi_1 (\omega-\omega_0) + \frac{\varphi_2}{2!} (\omega-\omega_0)^2 + \\
    \frac{\varphi_3}{3!} (\omega-\omega_0)^3 + \frac{\varphi_4}{4!} (\omega-\omega_0)^4 + \dots \label{EQN:SpecPhase}
\end{multline}

$\varphi_2$ and $\varphi_3$ are free parameters, and higher-order terms are ignored. SOD primarily controls the duration, while TOD controls the sharpness of the leading or trailing edge as well as the ringing in the field on the opposing edge. It is the interplay between SOD and TOD that then determines the pulse duration and the shape. By defining the ratio between TOD and SOD,

\begin{equation*}
    \alpha = \frac{\varphi_3 / ps^3}{\varphi_2/ps^2},
\end{equation*}

we gain a single parameter to describe the general shape of a shaped pulse that is approximately invariant to pulse duration.

Based on the aforementioned LCLS-II photoinjector commissioning requirements~\cite{PhysRevAccelBeams.24.073401}, we focus on a nominal magnitude for SOD around 3.5 ps$^2$ so the upconverted pulse is $\approx$25~ps in time. Our fundamental pulse is modeled as a 1030 nm, 330 fs Gaussian with 50 uJ of energy after the Ytterbium-based front end for LCLS-II. To achieve a flat-top profile, $\alpha$  is initialized around $0.125$~ps. Once the spectral phase is applied, propagation and frequency mixing is handled using a symmetrized split-step Fourier method along with a fourth-order Runga-Kutta algorithm to solve the coupled nonlinear equations~\cite{boyd2020nonlinear}. In the {first nonlinear crystal, the thickness is 2 mm and the} crossing angle is set to 1.5~deg such that the sum-frequency signal could be separated from the two input beams and suppress intra-beam SGH but not so large as to reduce the spatial overlap of the beams in the crystal~\cite{raoult1998efficient}.

\begin{figure}
    \includegraphics[width=0.95\linewidth]{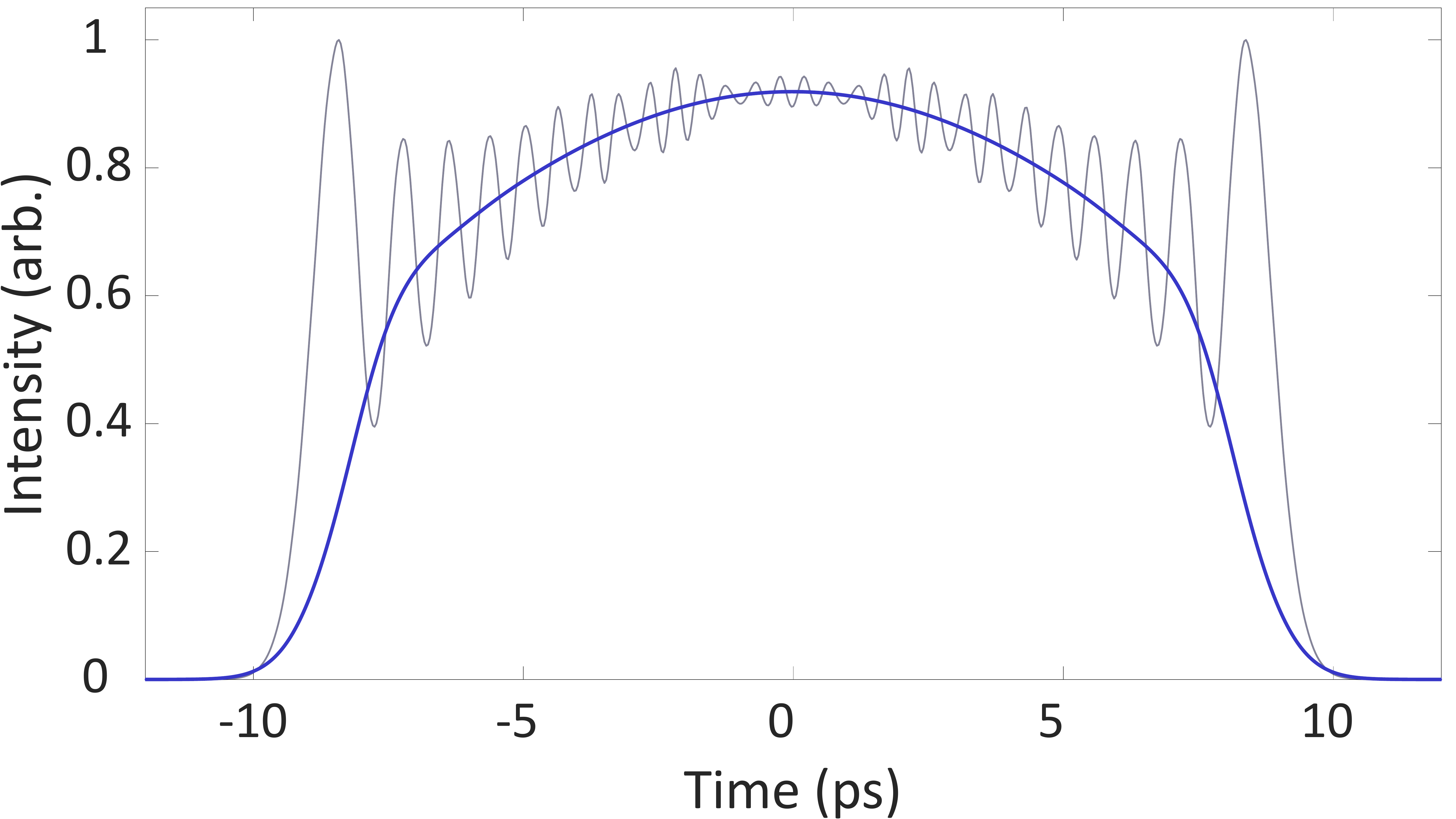}
    \caption{Temporal intensity profile of the SFG DCNS pulse directly after generation (grey) and the same pulse after a super-Gaussian spectral filter (blue).}
    \label{fig:GA_Pulse}
\end{figure}

The resultant pulse (Fig. \ref{fig:GA_Pulse}) displays the attractive qualities of a sharp rise time and a flatter profile than the traditionally used Gaussian pulses with upwards of 40\% conversion efficiency directly to the SFG signal in simulation. However, it is also characterized by large and rapid amplitude fluctuations (Fig. \ref{fig:GA_Pulse}, grey) on the picosecond scale that can be detrimental to electron beam emittance. These fluctuations can be remedied by applying a spectral amplitude filter after conversion that is significantly wider than the FWHM of the pulse's spectral bandwidth. With this, the interference from the high-frequency components can be attenuated without a major efficiency penalty. In the simulation, this is modeled as a second-order super-Gaussian spectral filter with $1/e$ width of 0.5 nm. Filters with performance appropriate for this can be found from commercial suppliers in interference or Bragg grating designs. After filtering, the total power in the field is reduced by less than 10\% and results in a smoother temporal profile (Fig. \ref{fig:GA_Pulse}, blue).

To generate the UV pulse profile for driving the cathode, the intensity profile of the smoothed SFG pulse is directly squared approximating the nonlinear conversion process. This has been shown to result in less than a 1\% RMS error between a full conversion simulation in crystal\cite{lemons2022temporal} while allowing for a significant computational speed-up useful to our optimization studies. Due to the flat phase structure and highly narrowband spectral content of the SFG pulse, conversion efficiency to the UV can be remarkably high. Full simulations of the UV conversion step, including the prerequisite phase modification and SFG conversion stages, result in pulses with 6 uJ of energy.

\section{Injector Simulations}\label{sec:inj}
\begin{figure*}
    \centering
    \includegraphics[width=\linewidth]{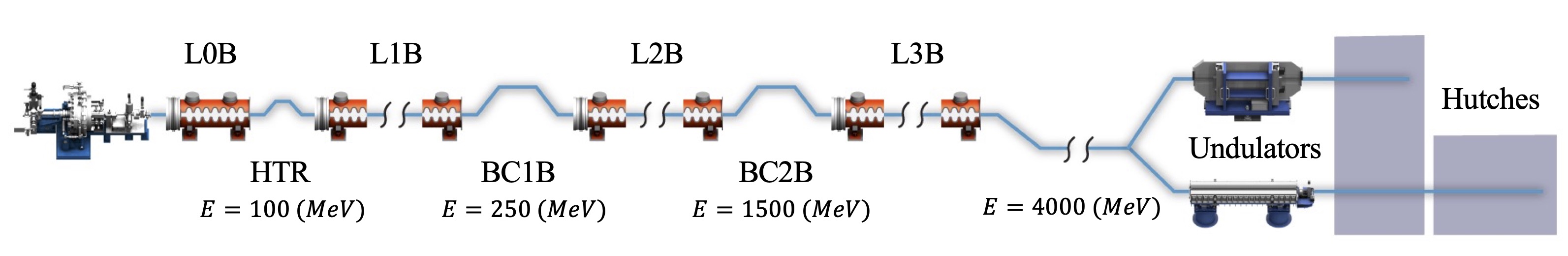}
    \caption{LCLS-II injector and linac layout, where L0B is the injector, L1B-L3B are the linac areas, HTR is the laser heater, and BC1B and BC2B are bunch compressors. }
    \label{fig:lcls-II}
\end{figure*}

The LCLS-II photoinjector consists of a 187 MHz quarter wave RF gun, followed by a two cell 1.3 GHz buncher, two solenoids, and several quad and correctors. The layout of the photoinjector and commissioning details can be found in~\cite{PhysRevAccelBeams.24.073401}. The beam is then injected into a cryomodule consisting of eight superconducting niobium cavities. The beam reaches an energy of 100 MeV when exiting the cryomodule at approximately 15 meters from the cathode. This portion of the machine will be referred to as the injector for the remainder of this paper, and simulations shown in this section take space charge into account. The same bunch charge, 100~pC, is used for both Gaussian and DCNS simulations. Following the inejctor is the remaining cryomodules and two bunch compressors as shown in Fig.~\ref{fig:lcls-II}.

Optimization studies of LCLS-II are ongoing in support of commissioning goals and expectations. A typical optimization consist of ten or more variables and two objectives. The variables used for this study include variables normally used for commissioning studies such as solenoids, cavity gradients, and cavity phases. Two additional variables were added to adjust the DCNS pulses: pulse duration (PD) and a scale factor. The PD effectively controls the FWHM of the UV pulse on the cathode while the scale factor is a multiplier to $\alpha$ and controls the `squareness' of the pulse, as described by the second and third order dispersion discussion in Section~\ref{sec:optics}. All variables and ranges for this simulation study can be found in Table~\ref{tab:params}.\footnote{We note that since the completion of this study, it has been determined that the buncher gradient will nominally be fixed at 1.8 MV/m. }
\begin{table}\centering
	\begin{tabular}{@{}lccl@{}}\toprule
		Variable                   & Minimum & Maximum & Unit \\ \midrule
        Laser radius              & 0.15      & 0.75      & mm\\
        {PD}       & 1.0       &   30      & ps   \\
        Scale factor              & 0.0       & 1.25     & arb. \\
		Gun phase                 & -20.0     & 10.0     & Degrees\\
		Solenoid strength(s)     & 0.02       & 0.07     & T/m \\
		Buncher gradient         &  1.0       & 3.5      & MV/m \\
		Buncher phase            & -100       & -10.0    & Degrees \\
		Cavity gradient 1-4      & 0.0 		  & 32.0     & MV/m \\
		Cavity phase 1-4        & -40.0       & 40.0      & Degrees \\
	\end{tabular}
	\caption{Variables and boundaries used in simulation optimizations of the LCLS-II superconducting injector.}
 \label{tab:params}
\end{table}

\begin{figure*}[htb]
\subfloat[\label{scatter_subfig:a}]{%
  \includegraphics[width=0.95\columnwidth]{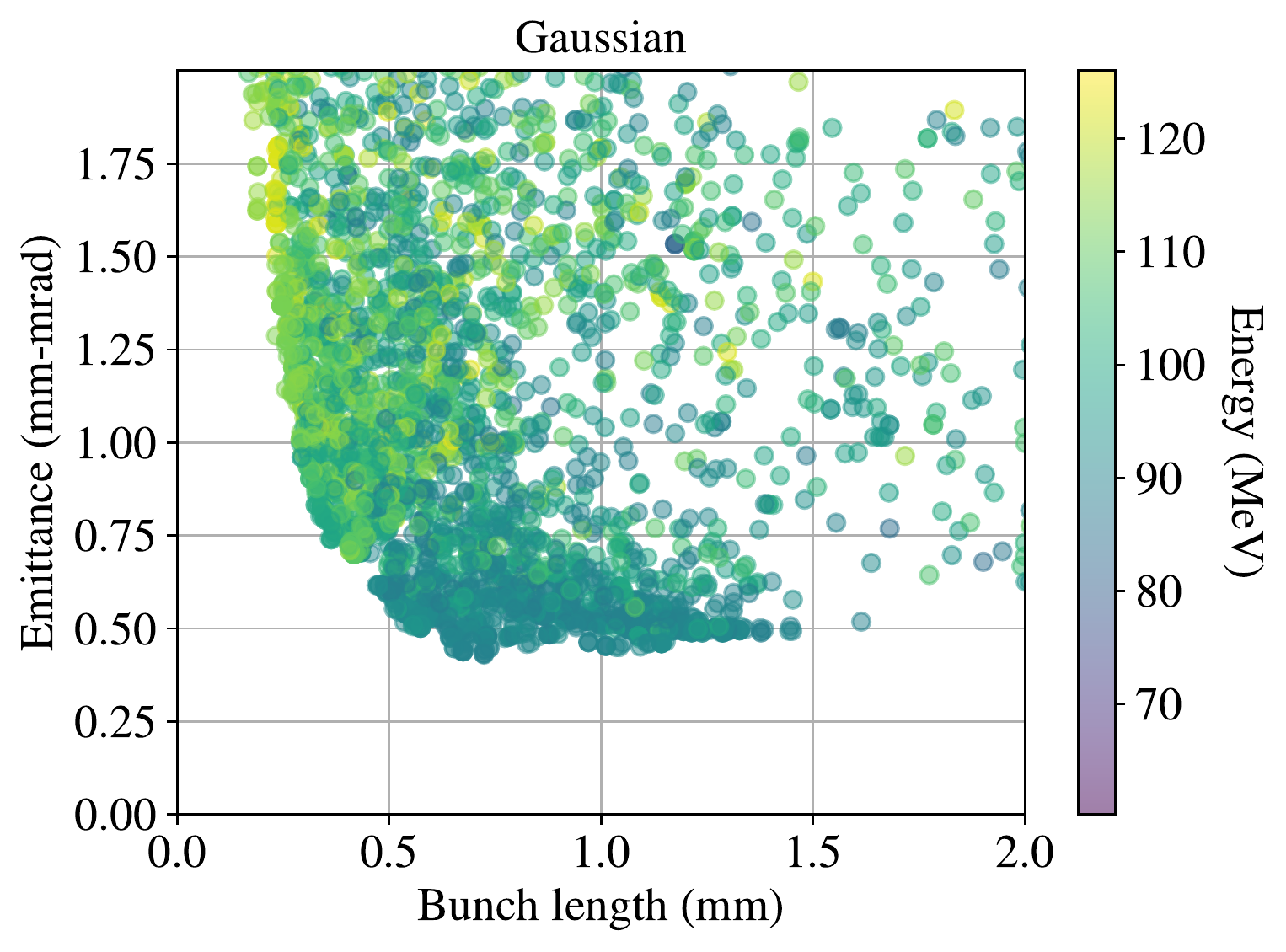}%
}\hfill
\subfloat[\label{scatter_subfig:b}]{%
  \includegraphics[width=0.95\columnwidth]{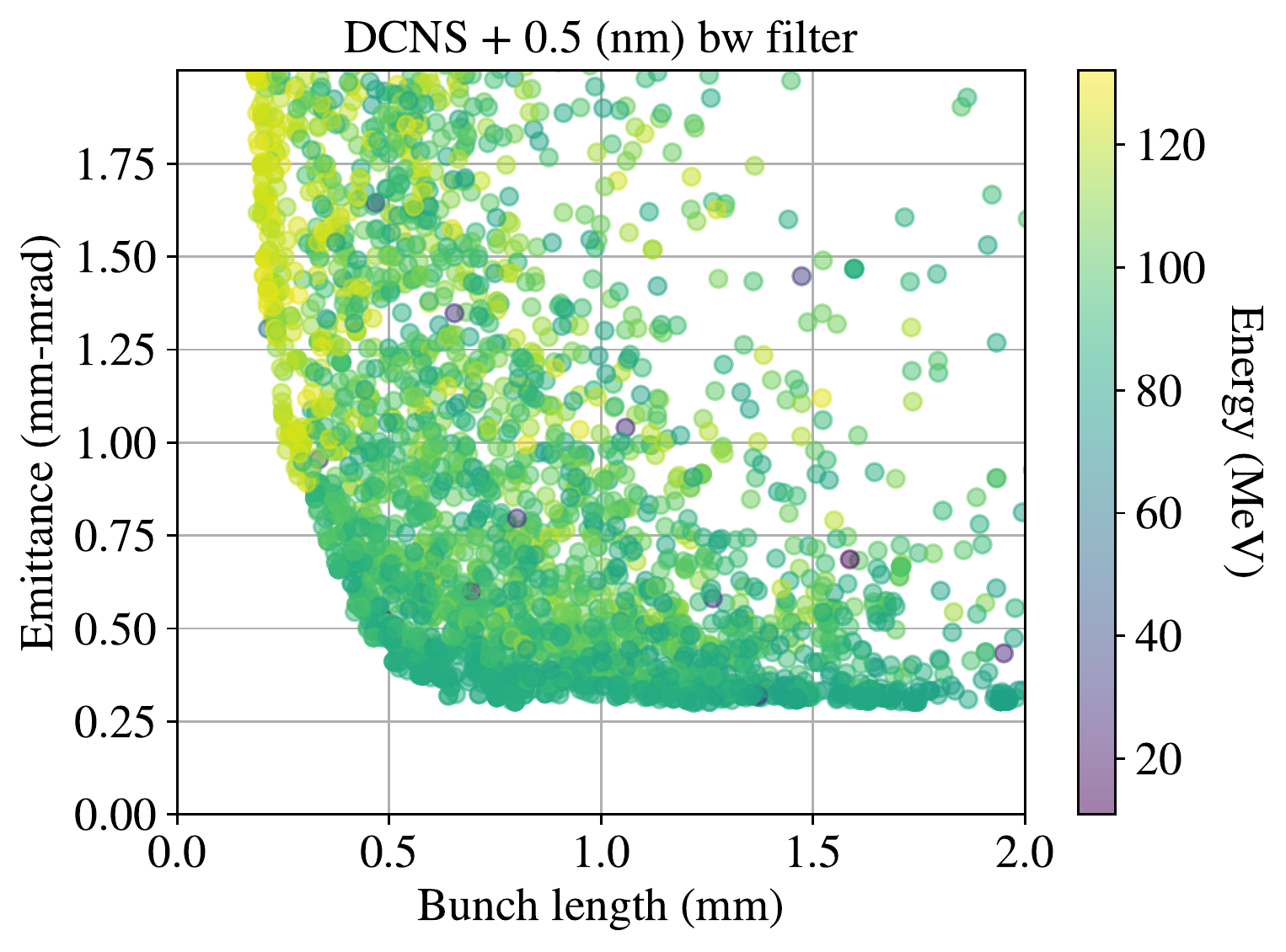}%
}
\caption{Scatter plots of Gaussian~(a) and DCNS simulation results~(b) in the objective space at the end of the injector (L0B). Three filter bandwidths (bw) were simulated, but only the best result is shown here (0.5~nm). To emphasize the expected operating region, the full emittance and bunch length range is not shown. Simulation points with emittance and bunch length values larger than 2.0 (mm-mmrad/mm) are not shown here. The color bar represents the mean energy of a given simulation. Lighter regions of the plot indicate simulations with larger mean energy values (i.e. 100 MeV and up).}
\label{fig:scatter}
\end{figure*}

The objectives of interests for this case are the emittance ($\epsilon_x$) and bunch length ($\sigma_z$) at the end of the injector (15 meters). Only the emittance in the x dimension is considered when determining the transverse performance, as 2D symmetric field maps are used. The popular algorithm NSGA-II~\cite{dpam:02,PhysRevSTAB.16.010101} was used for optimization runs in combination with the PIC simulation code OPAL~\cite{2019arXiv190506654A}. Four optimization cases were simulated using the variable ranges in Table~\ref{tab:params}, or in the case of the Gaussian optimization, a subset of those variables (excluding PD and scale factor). Three filter bandwidths were simulated: 0.5, 0.7, and 1 nm. Figure~\ref{fig:scatter}, shows the resulting Pareto points for the Gaussian and the DCNS case with the lowest emittance.

From Fig.~\ref{fig:scatter}, Gaussian and DCNS performance can be compared for strictly better points in the Pareto front. For nearly all bunch lengths, the DCNS pulses achieve better emittance values when compared to the Gaussian pulses. While this shows promise for the DCNS technique, it does not completely translate to performance expectations. For a practical comparison, in the case of LCLS-II, we compare Gaussian vs. DCNS pulse performance at the 1 mm bunch length target for operations. At the 1~mm bunch length in Fig.~\ref{fig:scatter}, the Gaussian pulses do not reach an emittance below 0.4 $\mu$m. For all filter widths, DCNS pulses are able to produce beams with emttances below 0.4 $\mu$m with a 1~mm bunch length.

\begin{figure*}[htb]
\subfloat[\label{injectorCurrent_subfig:a}]{%
  \includegraphics[width=0.95\columnwidth]{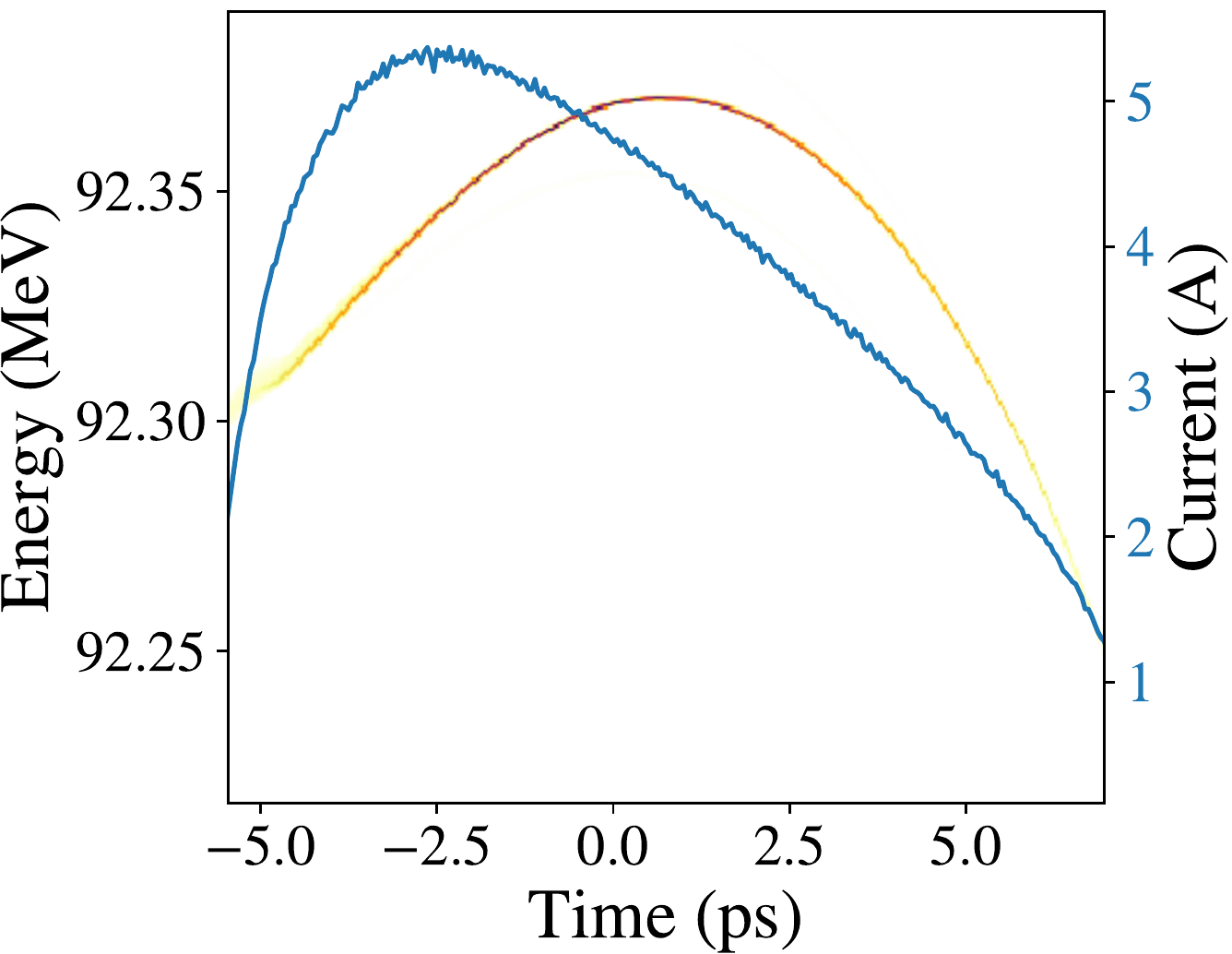}%
}\hfill
\subfloat[\label{injectorCurrent_subfig:b}]{%
  \includegraphics[width=0.95\columnwidth]{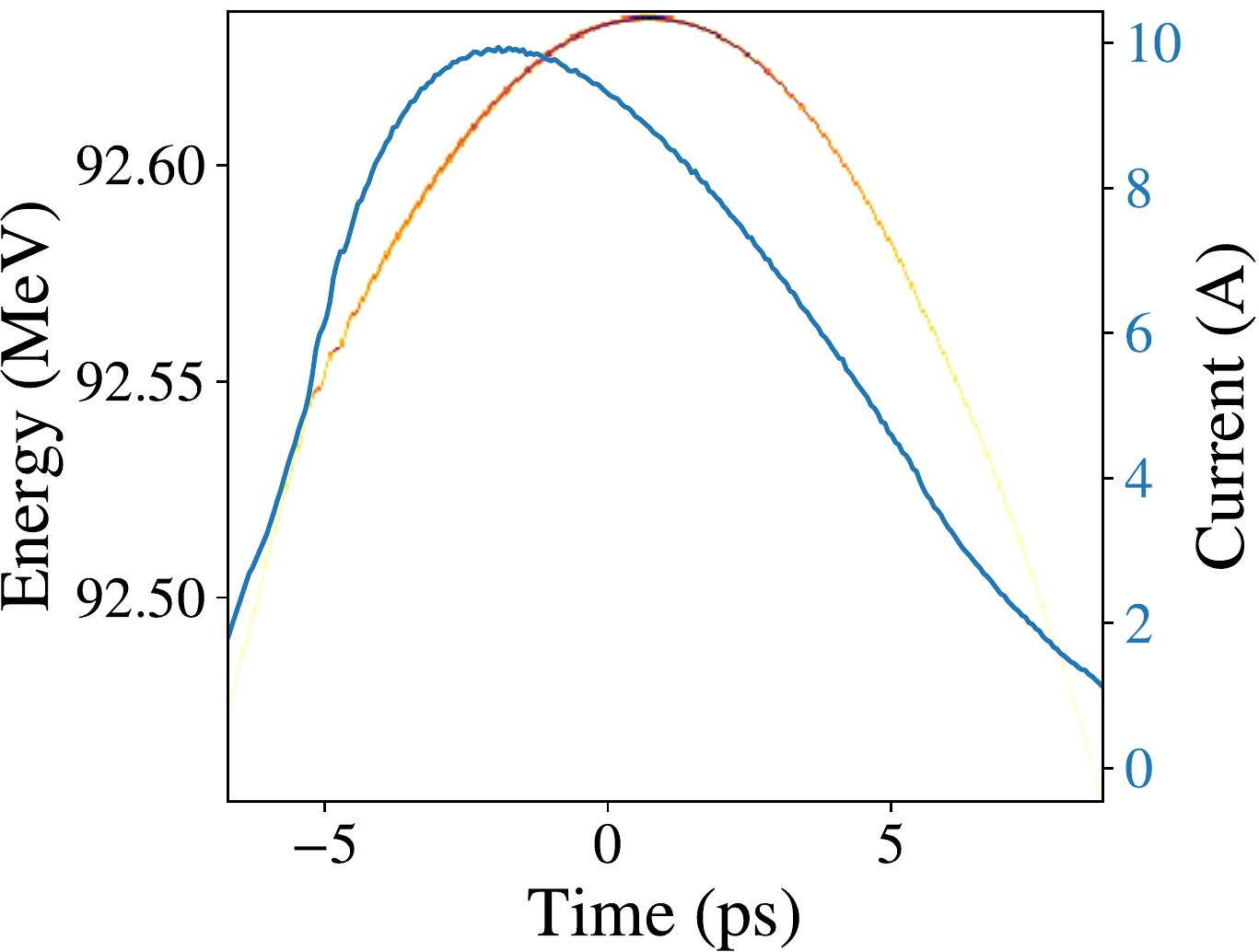}%
}
\caption{Energy spread and current distribution at the end of the injector~(L0B) for the selected Gaussian~(a) and DCNS~(b) pulse. These beam distributions are located after L0B, and before the entrance to the laser heater. The DCNS point was chosen from the optimization run using a 0.5~nm bandwith filter.}
\label{fig:injectorCurrent}
\end{figure*}

For further comparison, the best emittance DNCS pulse at 1 mm bunch length is chosen to send downstream. The comparison point was chosen from the 0.5 nm bandwidth DCNS case, with an emittance value of approximately 0.37$\mu$m. Before passing this beam on to linac simulations, the fidelity was increased from 50,000 to 10 million particles and the the chirp was adjusted. Non-ideal chirp is observed in normal Gaussian optimizations as well; it is not a result of using the DCNS pulses on the cathode. A small phase scan in cryomodule cavity four is done to reduce the chirp at the end of the injector. At this location, roughly 7.5 m, the emittance is sufficiently compensated, and a slight change in the fourth cavity phase will not dramatically change the emittance value. However, adjustment to the phase here does impact the energy spread enough to improve bunch compression down stream. The final energy spread and current distribution handed off to linac simulations are shown in Fig.~\ref{fig:injectorCurrent}.

\section{Linac Simulations}\label{sec:linac}
The LCLS-II layout has a total 35 cryomodules consisting of eight 1.3 GHz niobium cavities each. The first cryomodule is followed by a laser heater section. Two high frequency (3.9 GHz) cavities are installed after the third cryomodule, and the high frequency cavities are followed by a bunch compressor (BC1). The second bunch compressor follows cryomodule 15. The current LCLS-II layout is shown in Fig.~\ref{fig:lcls-II}.

Optimization of the LCLS-II linac for commissioning has already taken place at 20~pC, 50~pC, and 100~pC~\cite{technote2} with Gaussian-shaped UV pulses at the cathode. The linac phases and bunch compressor settings were adjusted to minimize energy spread while maintaining emittance. Further optimizations were done to increase the peak current of the Gaussian beams to 2 kA. This work was used as a starting point for comparison to the DCNS pulses.

\begin{figure*}[htb]
\subfloat[\label{preUndCurrent_subfig:a}]{%
  \includegraphics[width=0.95\columnwidth]{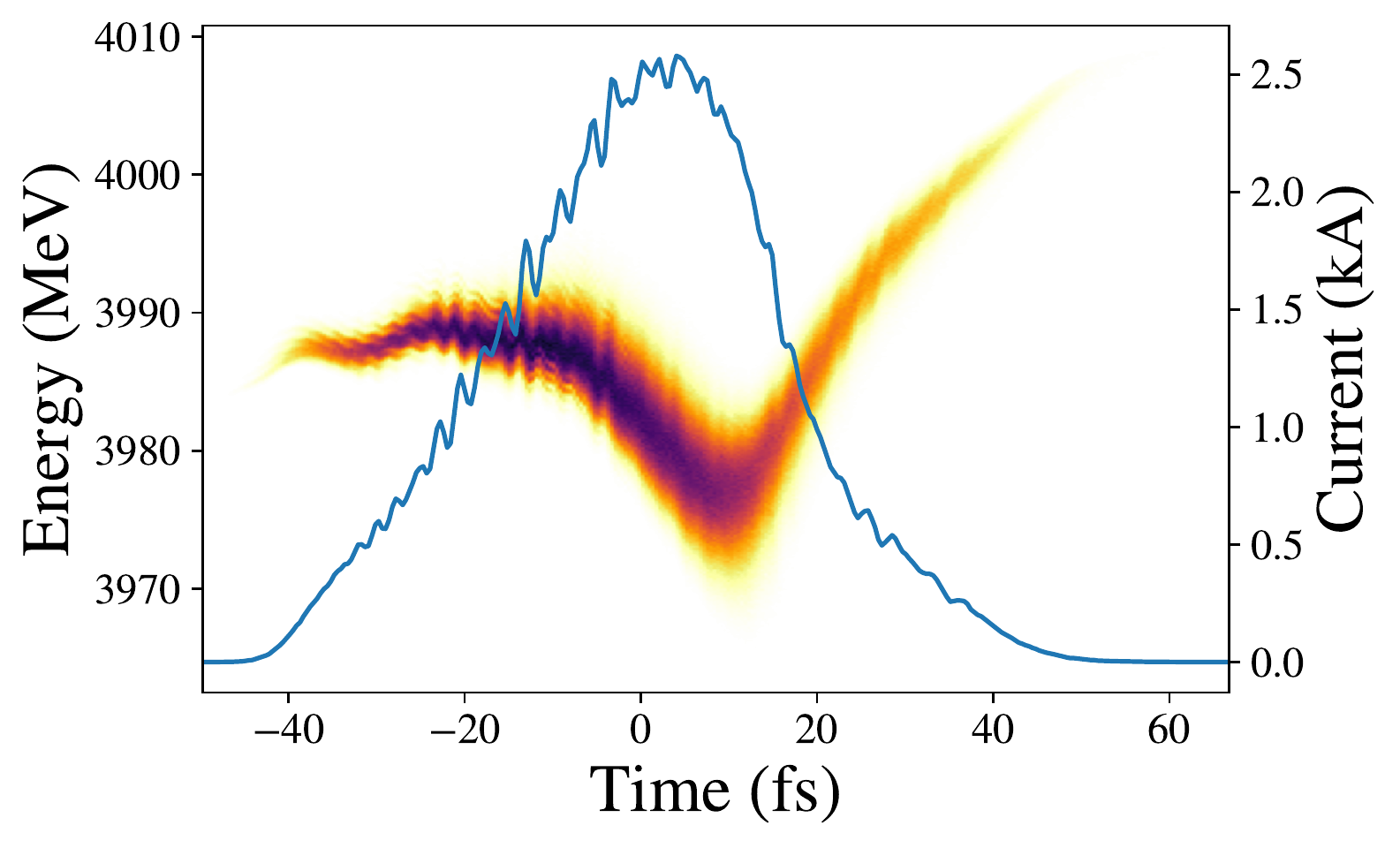}%
}\hfill
\subfloat[\label{preUndCurrent_subfig:b}]{%
  \includegraphics[width=0.95\columnwidth]{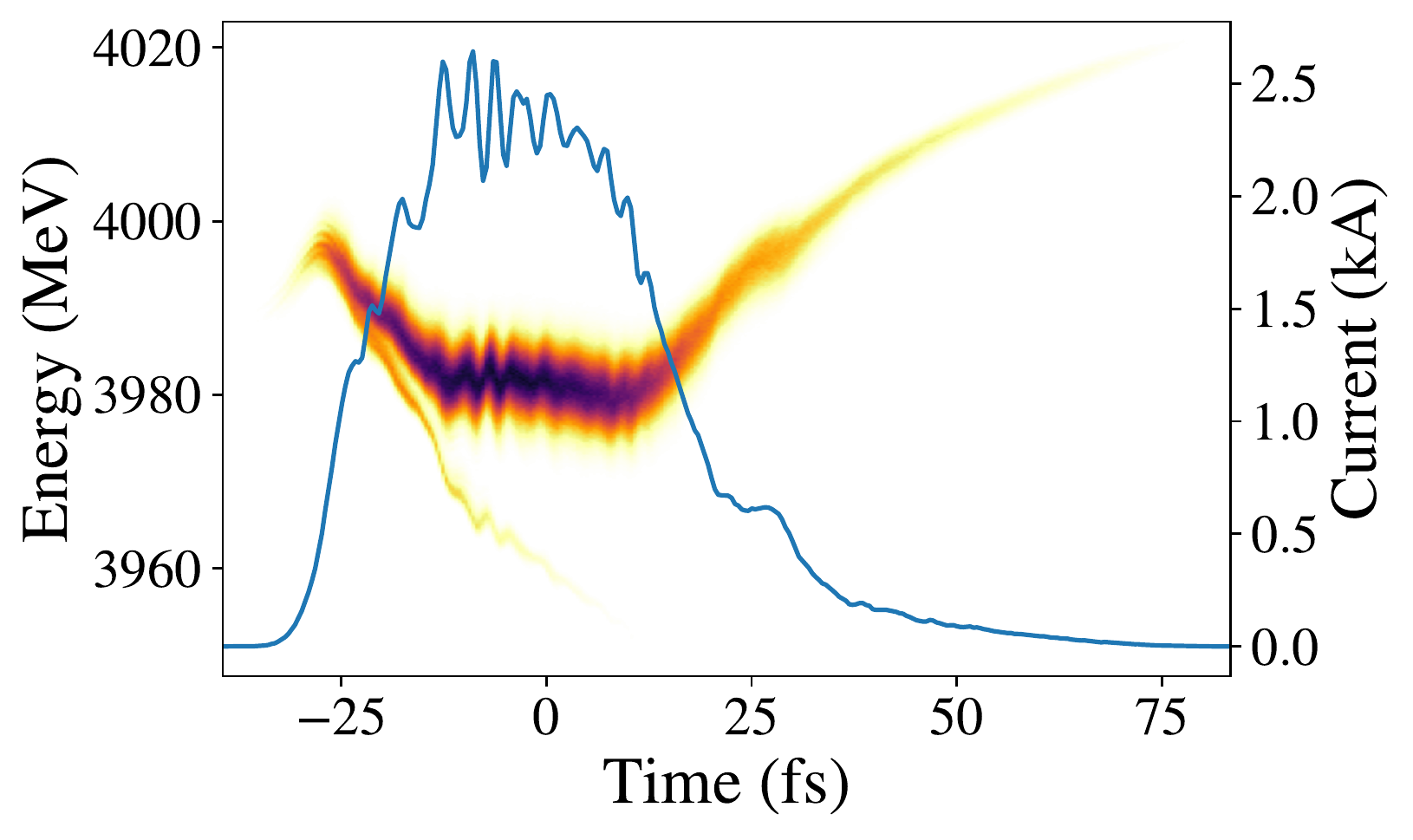}%
}
\caption{Gaussian beam with peak current of 2.1 kA (a), and DCNS beam with peak current of 2.2 kA (b) at the end of the linac~(L3B). Both bunches were taken from linac optimization of LCLS-II. These beam distributions are located before the undulator lines.}
\label{fig:linac_compare_gauss_dcns}
\end{figure*}

Simulation of both the Gaussian and DCNS beams in the linac
used the code Elegant. The beam distributions shown in Fig.~\ref{fig:injectorCurrent} were propagated from the end of the first cryomodule to the start of the soft x-ray line. For initial comparison, the same linac settings were used for both the Gaussian and DCNS pulses. One would expect some adjustment needed when switching initial conditions into the linac; however, the DCNS pulse performed well with little adjustment. Compared to the Gaussian pulse, the DCNS beam reaches higher peak currents for a larger portion of the beam, see Fig.~\ref{fig:linac_compare_gauss_dcns}. This comparison suggests a larger portion of the DCNS beam will lase, which we later confirm after analysis of the slice emittance in Section~\ref{sec:fel}.

To confirm the DCNS pulse results in the linac, an additional round of optimization was done using several parameters downstream of the injector, including linac phases (L1B, L2B), microbunching compensation chicanes (BC1B, BC2B), and laser heater power and waist (cross section based on rms).
The objectives were to minimize energy spread while maximizing the peak current at the soft x-ray undulator entrance.

In this case, the objective was not a scalar value as in Section~\ref{sec:inj}. In order to calculate the peak current, a time window of the bunch was taken over 70\% of the charge. This time window was shifted from beginning to end of the bunch and the average current was calculated for the 70\% window for each time delay. The window with the maximum current was chosen for comparison to other simulations. This process was then repeated for finding the minimum rms slice energy spread. Finally, this process was repeated a third time to find the window that maximizes the metric of avg current divided by rms energy spread. The resulting Pareto front is shown in Fig.~\ref{fig:linacsimresults}. These optimization results recovered the initial results seen without optimization, i.e. DCNS out performing Gaussian with the same linac parameters. In addition to improved performance at nominal linac parameters, the optimizations indicating much larger peak currents could be achieved. However, the price to achieve the larger peak currents comes at larger emittances and non-linear phase spaces at the end of the linac.
\setcounter{figure}{7}
\begin{figure}
    \centering
    \includegraphics[width=0.95\linewidth]{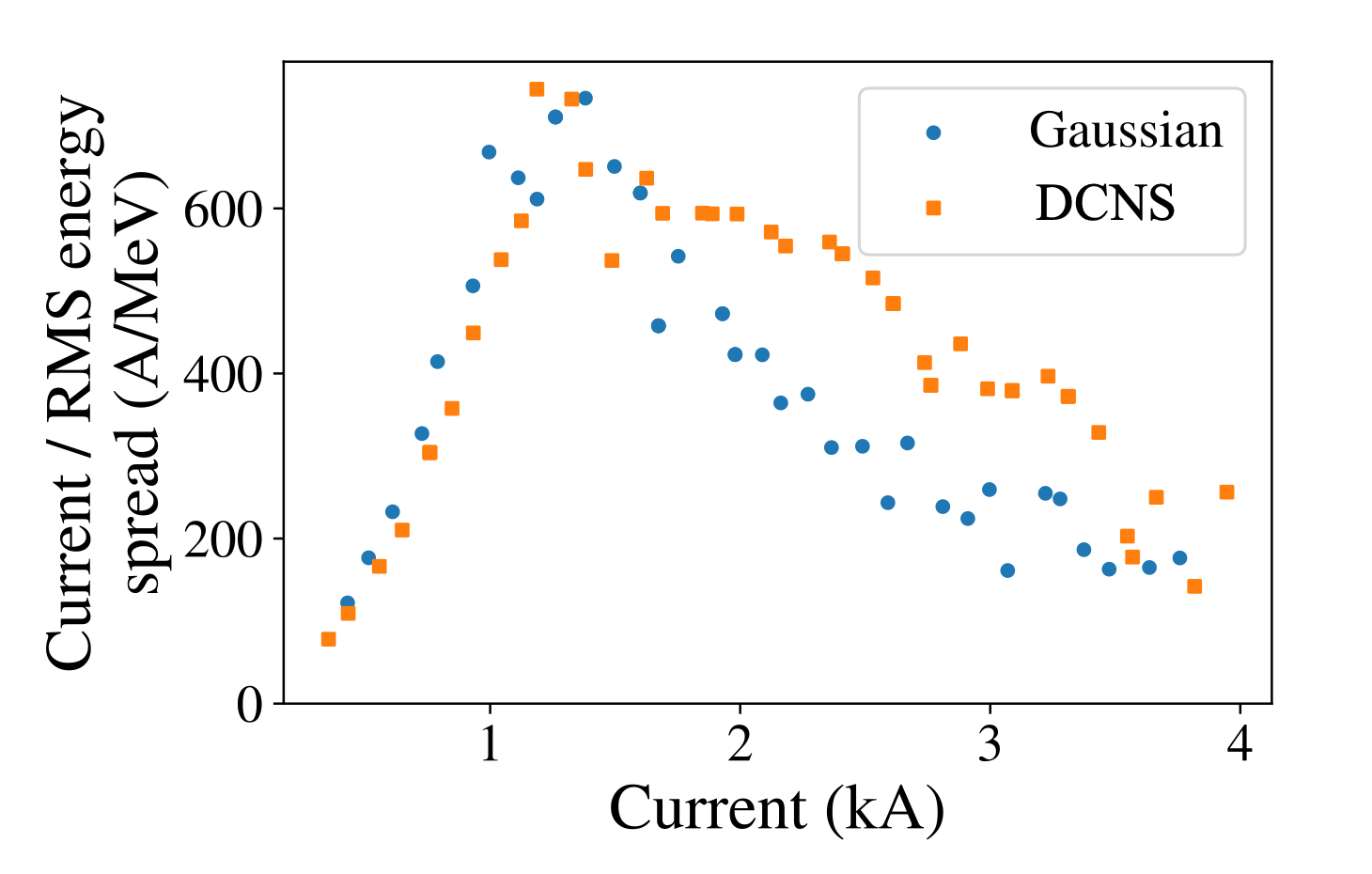}
    \caption{Resulting Pareto front from Gaussian and DCNS optimization of the LCLS-II linac.}
    \label{fig:linacsimresults}
\end{figure}

\section{FEL Estimates and Comparisons}\label{sec:fel}
Before comparison of the two bunches simulated in Section~\ref{sec:linac}, we consider the current frame work of LCLS-II and LCLS (normal conducting accelerator) performance metrics. The undulator gap can vary from 7.2 mm to 100~mm (essentially infinite distance with respect to the beam), with 7.2~mm-15~mm being the nominal range~\cite{undparams}. Pulses with 3 mJ of x-ray energy per pulse are considered excellent performance on LCLS.
Peak currents of up to 4.5~kA are possible on the LCLS (copper) to hard x-ray line, and conservative LCLS-II commissioning estimates predict 1~kA on LCLS-II superconducting to soft x-ray line (SXR). A planned upgrade to LCLS-II, LCLS-II-HE, will boost the beam to 8~GeV and 2.0~kA peak current.

\begin{figure*}[htb]
\subfloat[\label{preUnsubfig:a}]{%
  \includegraphics[width=0.95\columnwidth]{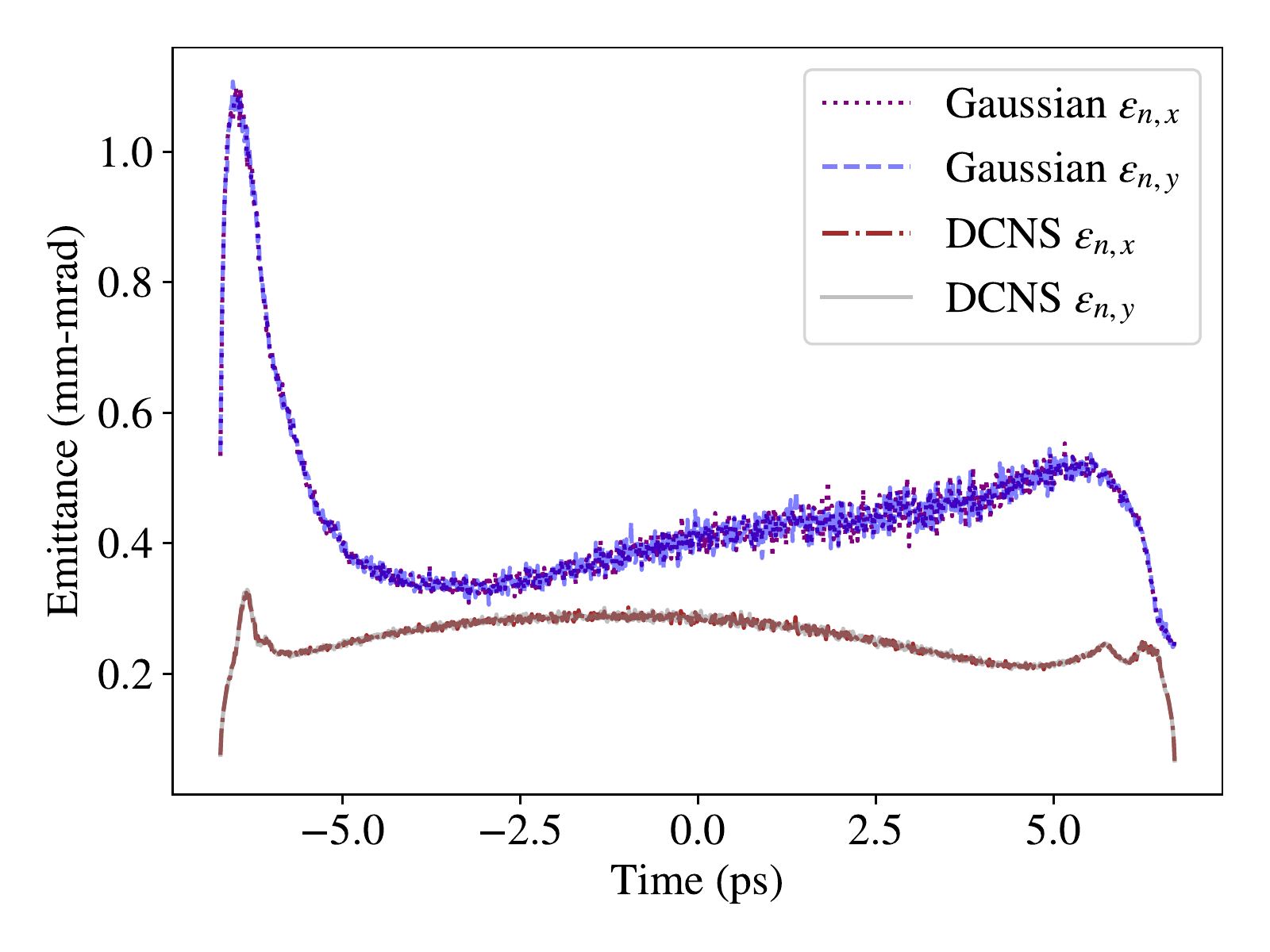}%
}\hfill
\subfloat[\label{subfig:b}]{%
  \includegraphics[width=0.95\columnwidth]{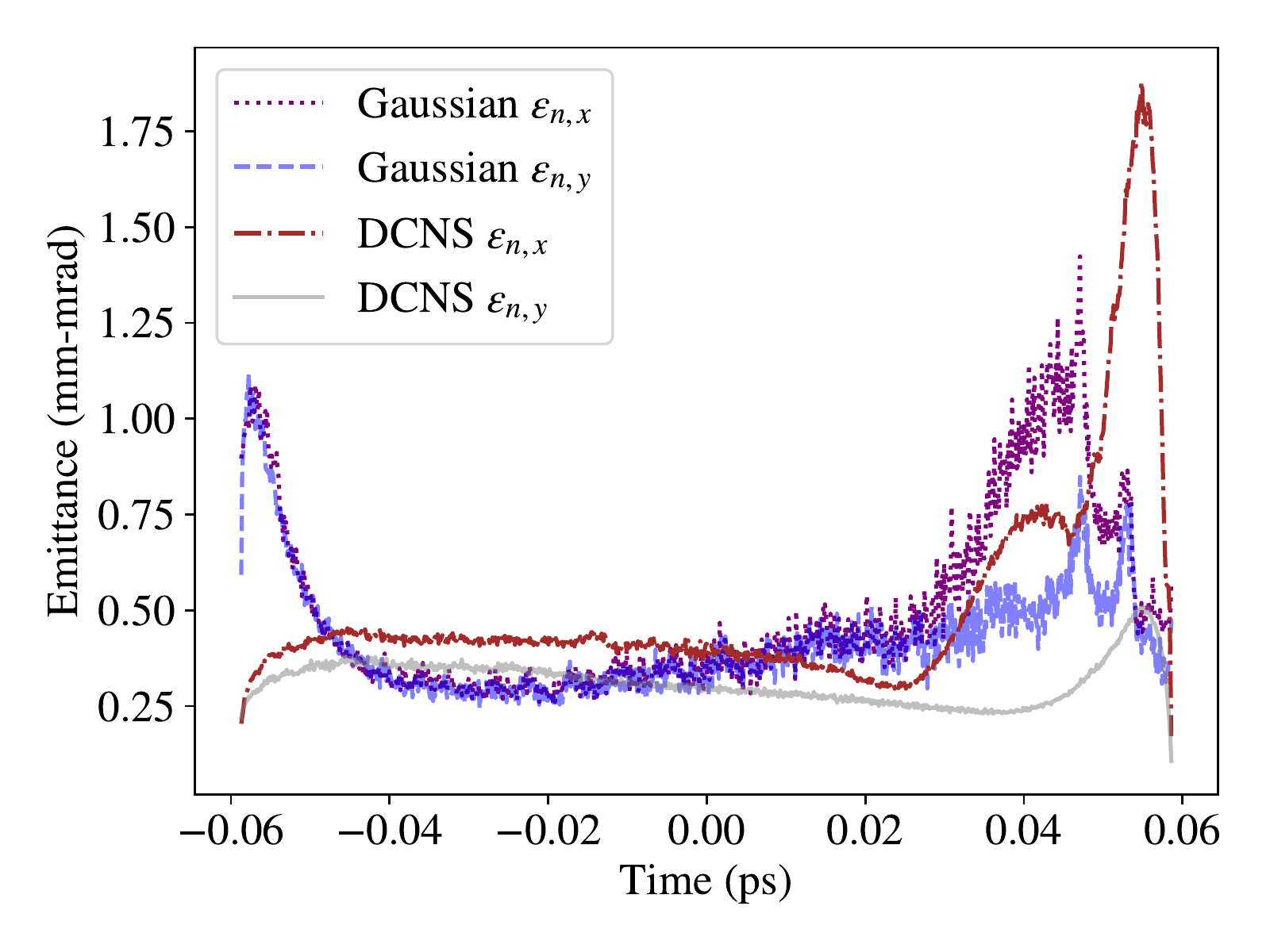}%
}
\caption{Slice emittances plotted at the end of the injector~(a) and start of the soft x-ray undulator line~(b). Both the Gaussian and DCNS simulation results reach low emittance values in the core.}
\label{fig:slicelinac}
\end{figure*}

In Fig.~\ref{fig:linac_compare_gauss_dcns}, both the Gaussian and DCNS simulations show the potential for a peak current greater than 2.0~kA, double the conservative LCLS-II estimates. However, peak current alone is not enough to guarantee good x-ray performance, the slice emittance over the peak current regime must be low enough to ensure lasing. From the slice emittance plots in Fig.~\ref{fig:slicelinac}, it appears the cores of both the Gaussian and DCNS pulses reach acceptable slice emittance values. Note, the distributions in Fig.~\ref{fig:slicelinac}a are symmetric because the beam has not passed through any bending or compression elements at this point.

Notable differences between the two cases appear in the head and tail of the bunches after the linac (Fig.~\ref{fig:slicelinac}), where the slice emittance more than doubles compared to the core. The core part of the DCNS beam at high current peak is very flat, i.e. less energy chirp than the conventional Gaussian beam. To quantify the percentage of each beam that is likely to contribute to lasing, slices below 0.5 um emittance were counted. In the Gaussan case, about 68.8~\% of slices in x and 79.8~\% of slices in y had emittance values below 0.5 mm-mrad. For the DCNS case, 78.1~\% and 99.3~\% of slices in x and y had emittance values below 0.5 mm-mrad. For the following FEL estimates, the average of the two, 74.3~\% for Gaussian and 88.7~\% for DCNS, were used to estimate how much charge contributes to lasing. With a total simulated bunch charge of 100~pC, this gives about 74~pC and 89~pC for the Gaussian and DCNS bunches respectively. Similarly, when calculating the energy and emittance values used in the following FEL power estimates and table, the average was taken over the core of the bunch. In this paper, the core of the bunch is defined as the portion of the beam with emittance values below 0.5~mm-mrad. In this way, charge in the large emittance tails is not considered to contribute positively to the FEL estimates.
\begin{table}\centering
    \begin{tabular}{@{}lccl@{}}\toprule
        Variable        & Gaussian & DCNS  & Unit \\ \midrule
        Charge          & 74.0     & 89    & pC   \\
        Energy          & 4.0      & 4.0   & GeV  \\
        Energy spread   & 0.005    & 0.01 & $\frac{dE}{E}$ \\
        Average Current & 1.59     & 1.96  & kA \\
        Emittance       & 0.36     & 0.35  & mm-mrad \\
        Undulator K     &  3.5 & 3.5 & \\
        Undulator Length & 70  & 70 & m\\
    \end{tabular}
    \caption{Parameters used to calculate FEL (SASE) power estimate and coherent x-ray energy per pulse. Beam specific parameters were taken from linac simulation results. Beam parameters are taken from averages in the core, where emittance values are below 0.5 mm-mrad. Undulator parameters are taken from expected or demonstrated LCLS-II values~\cite{undparams}.}\label{tab:fel}
\end{table}
\setcounter{figure}{9}
\begin{figure}[!ht]
    \centering
    \includegraphics[width=0.95\linewidth]{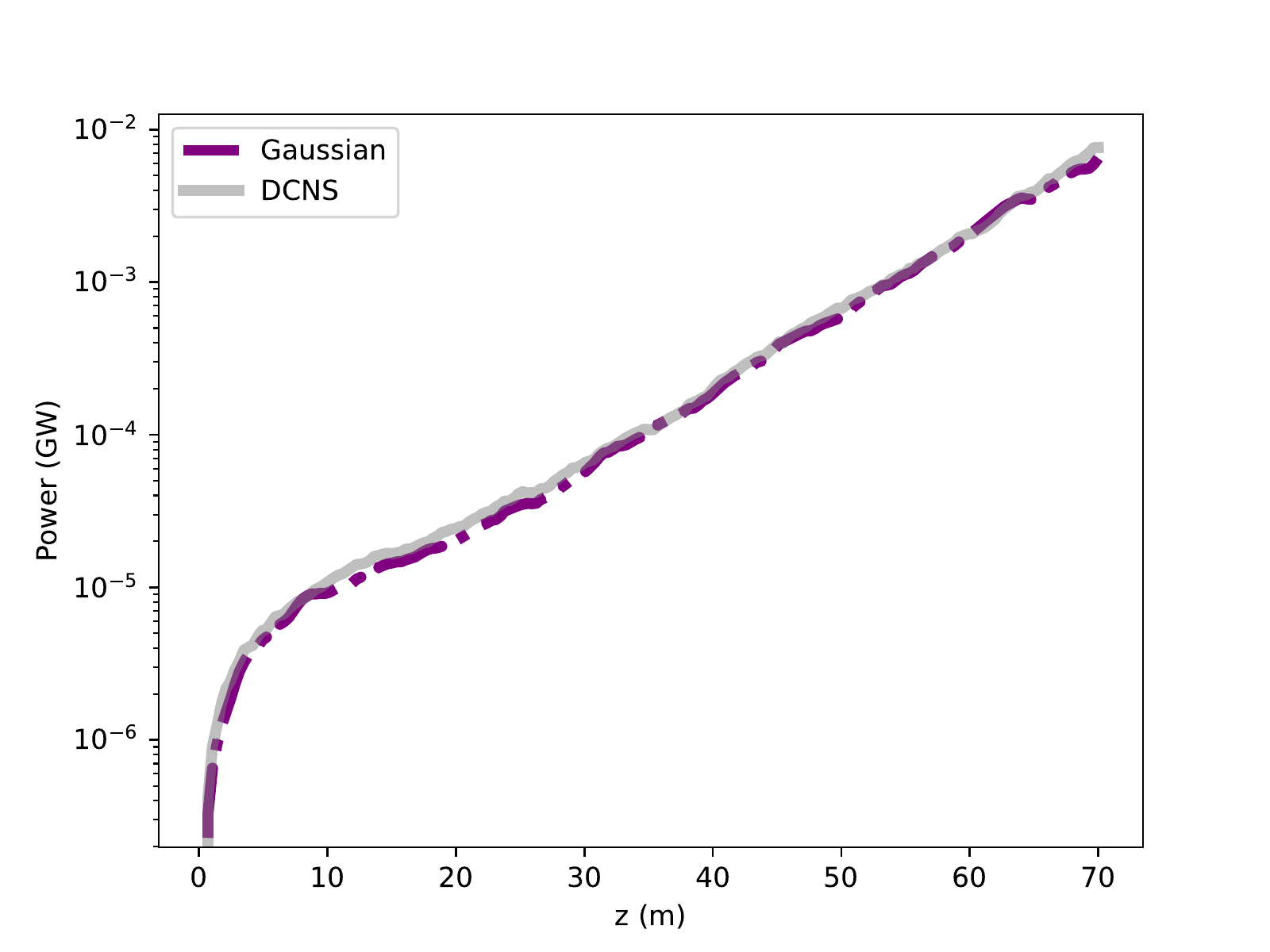}
    \caption{FEL (SASE) power estimate for both the Gaussian and DCNS beams. Neither case reaches saturation in 70~m.}
    \label{fig:felpower}
\end{figure}

With similar energy and core emittance values, no large differences in the SASE power calculation appear in Fig.~\ref{fig:felpower}, and both simulations do not reach saturation in 70~m. Next we estimate the coherent x-ray energy per pulse and take into consideration the difference in charge at low emittance. In Fig.~\ref{fig:photons}, it is assumed the undulator gap can be varied across the nominal range (7.2~mm to 15~mm) to generate photons of different energies given the same beam energy, emittance, and charge. Here is where the benefit of DCNS is clear, with larger x-ray energy per pulse values as shown in Fig.~\ref{fig:photons}, which resulted from slightly higher average current and lower emittance. Since the DCNS pulse maintains better emittance over more slices of the bunch, more of the bunch charge contributes to lasing. This results in the DCNS beam delivering higher x-ray pulse energy across the full available photon range. Note, we do not claim to have maximized FEL performance here, we only compared Gaussian and DCNS performance given nominal optimization of the injector and linac (similar peak current and energy spread ranges). Optimization of the DCNS shape and accelerator parameters could be formulated in a way that emphasizes FEL power as the main objective. Optimization of the DCNS pulse and accelerator parameters with updated objectives could lead to higher FEL powers. While an interesting problem worth investigating, we leave peak FEL power optimization for future study.
\setcounter{figure}{10}
\begin{figure}[!ht]
    \includegraphics[width=0.95\linewidth]{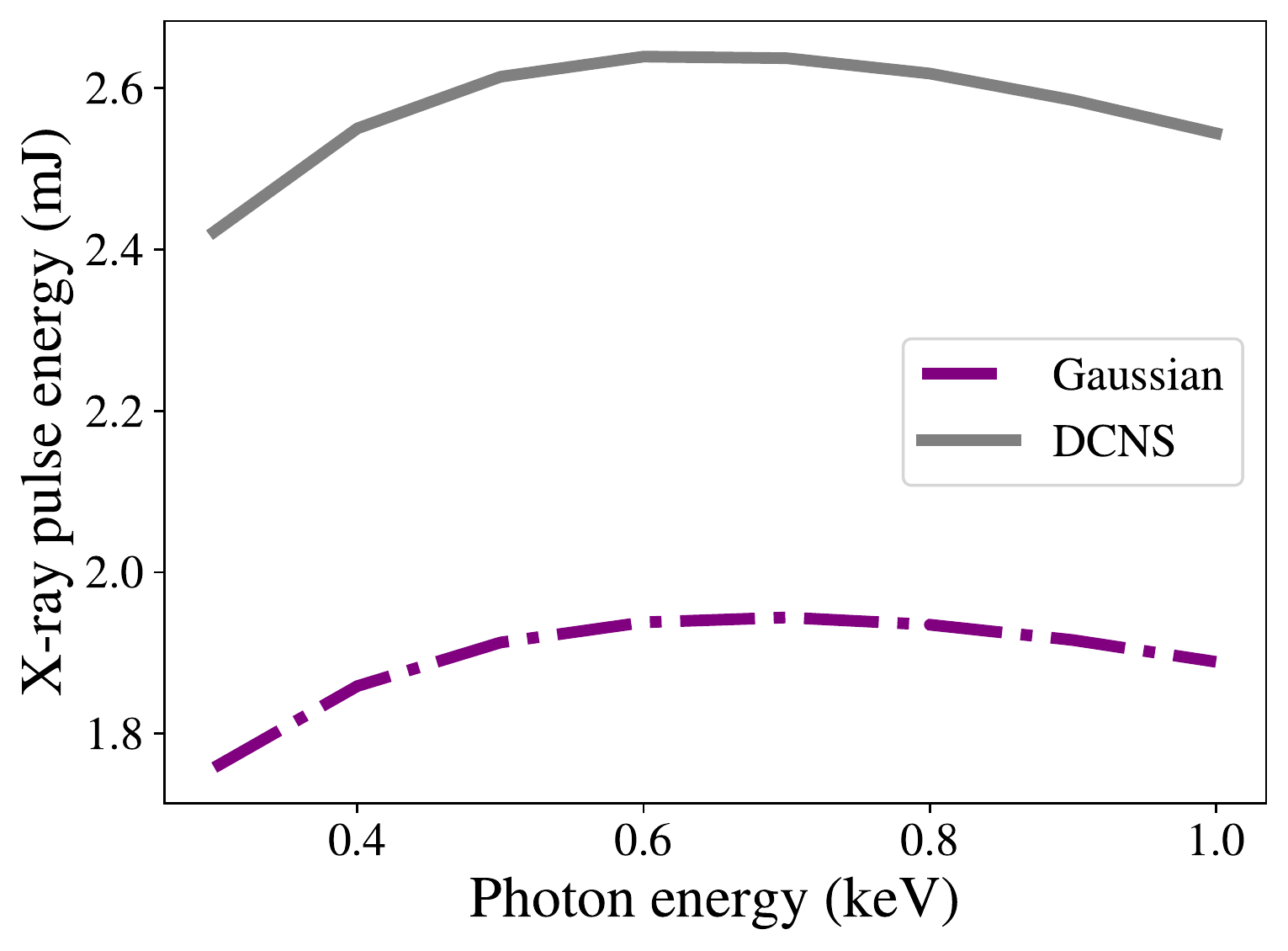}
    \caption{Photon energy compared to x-ray energy per pulse. Due to the higher peak current and low core emittance over more of the total charge, the DCNS pulse delivers 35\% more x-ray energy per pulse than the Gaussian case. }
    \label{fig:photons}
\end{figure}

\section{Conclusion}
In this paper, the UV laser shaping technique, Dispersion Controlled Nonlinear Shaping (DCNS), is reviewed and proposed for UV laser pulse shaping on photocathodes. Both the transverse and longitudinal shaping of UV pulses greatly impacts electron beam properties in photoinejctors and free-electron lasers. Gaussian and DCNS laser pulses were simulated then used to simulate laser pulses on the LCLS-II photocathode. The resulting e-beams were then simulated in the injector and linac of LCLS-II, which is currently being commissioned. Both the Gaussian and DCNS case, the laser pulse shapes were included as optimization variables along with magnets and phases for the injector and linac to ensure a fair comparison. Optimization results indicate DCNS pulses can reach lower emittances at the exit of the injector. The DCNS pulses were also able to reach higher peak currents in the linac.

Two optimized beam distributions, one Gaussian and one DCNS, were chosen for a basic FEL comparison. There is no appreciable difference in the time to saturation or SASE power generation. However, when the amount of low emittance charge in the core is taken into account, it was estimated that about 74 pC vs 89 pC out of 100 pC contributed to lasing in the Gaussian and DCNS cases respectively. Using the variable gap of the undulators to set photon energy, the potential x-ray energy per pulse in the DCNS case was estimated to be 35\% higher than the Gaussian case at ~2.6 mJ compared to ~1.9 mJ. Given the lower emittance values over a larger proportion of the beam, there may be potential to lower gain length in certain parameter regimes. If the gain length can be lowered through emittance manipulation, saturation could be reached in some regimes, resulting in easier extraction of x-rays. The flatness of the DCNS beam core (i.e. low chirp), could also contribute desired features in SASE photon beams, such as reduced bandwidth and higher spectral brightness.

In conclusion, DCNS is a promising method for emittance reduction and improved FEL performance. By updating optics, accelerator facilities may access lower emittance regimes than possible before. This fine longitudinal control can lead to better beam dynamics and in the case of LCLS-II, better FEL performance.

\section{Funding}
This work was funded by DOE contract Nos. DE-AC02-76SF00515 and DE-SC0022559.

\section{Acknowledgments}
We wish to acknowledge Greg Stewart for contribution of the LCLS-II layout figure. We thank Heinz-Dieter Nun for fruitful FEL discussions and guidance. We also thank the Scientific Data Facility at SLAC for providing computational resources that were used for this work.

\bibliographystyle{Frontiers-Vancouver}
\bibliography{frontiers}

\end{document}